\documentclass[epj]{svjour}
\usepackage{graphicx}
\usepackage{float}
\usepackage{rotating}

\newcommand{\be}{\begin{equation}}
\newcommand{\ee}{\end{equation}}
\newcommand{\ba}{\begin{eqnarray}}
\newcommand{\ea}{\end{eqnarray}}

\begin{document}
\title {The radiative decay of the $\Lambda(1405)$ and its two-pole structure}

\author{L. S. Geng\thanks{E-mail address: lsgeng@ific.uv.es}
 \and E. Oset\thanks{E-mail address: oset@ific.uv.es}
 \and M. D\"{o}ring\thanks{E-mail address: michael.doering@ific.uv.es}}

\institute{
Departamento de F\'{\i}sica Te\'orica and IFIC, Centro Mixto,
     Institutos de Investigaci\'on de Paterna - Universidad de Valencia-CSIC}

\abstract{
We evaluate theoretically the radiative decay widths into $\gamma\Lambda$
and $\gamma\Sigma^0$ of the two poles of the $\Lambda(1405)$ found in chiral unitary theories
and we find quite different results for each of the two poles. We show that, depending on
which reaction is used to measure the $\Lambda(1405)$ radiative decays, one gives more weight
to one or the other pole, resulting in quite different shapes in the $\gamma\Lambda(\Sigma^0)$
invariant mass distributions. Our results for the high-energy pole
agree with those of the empirical determination of the $\gamma\Lambda$ and $\gamma\Sigma^0$ radiative
widths (based on an isobar model fitting of the $K^-p$ atom data), which are sometimes referred to as
``experimental data''. We have made a detailed study of the $K^-p\rightarrow\pi^0\gamma\Lambda(\Sigma^0)$
and $\pi^-p\rightarrow K^0\gamma\Lambda(\Sigma^0)$ reactions and have shown that they, indeed, lead to
different shapes for the $\gamma\Lambda(\Sigma^0)$ invariant mass distributions.
} 

\maketitle

\section{Introduction}\label{sec:introduction}
The nature of the $\Lambda(1405)$ has been in dispute since the early days. In
most quark-model calculations, it is described as a $p$-state $q^3$ baryon with
mainly a SU(3) singlet structure~\cite{Isgur:1978xj}.  On the other hand, the $\Lambda(1405)$
resonance has long been suggested to be a bound state of the $\bar{K}N$ system,
and therefore of $q^4\bar{q}$ structure~\cite{Dalitz:1960du}.  In recent years,
this argument has been strengthened within the unitary extensions of chiral
perturbation theory
U$\chi$PT~\cite{Kaiser:1996js,Oset:1997it,Oset:2001cn,Oller:2000fj,Jido:2003cb,Garcia-Recio:2002td,Garcia-Recio:2003ks,Hyodo:2002pk}.
A particularly interesting discovery is that the nominal $\Lambda(1405)$ is a
superposition of two resonances. This was hinted in \cite{Oller:2000fj}
and studied in detail in \cite{Jido:2003cb}, where two poles were found
on the second Riemann sheet at $1390-i66$\,MeV and
$1426-i16$\,MeV, respectively. More recently, the studies of the $\bar{K}N$ interaction  have
been extended by including higher order chiral Lagrangians in the kernel of
the interaction~\cite{Borasoy:2005ie,Oller:2005ig,Oller:2006jw,Borasoy:2006sr}.
The position of the high-energy pole is rather similar in all these works and in
\cite{Jido:2003cb}, but there are variations in the position of the low-energy
pole. Nevertheless, the theoretical uncertainties have been studied in \cite{Borasoy:2006sr}
and the results of Ref.~\cite{Jido:2003cb} fit well within them.
As first demonstrated in Ref.~\cite{Jido:2003cb}, due to the
fact that these two poles couple differently to the coupled channels, different
reactions could observe different invariant mass distributions, thus offering
the possibility to experimentally test the two-pole prediction. The reactions $\gamma
p\rightarrow K^+\Lambda(1405)$ and $K^-p\rightarrow \Lambda(1405)\gamma$ are
shown to be sensitive to the high-energy pole of the $\Lambda(1405)$ and thus the corresponding
invariant mass distributions
exhibit a peak at $\sim 1420$\,MeV~\cite{Nacher:1998mi,Nacher:1999ni}. On the other hand, the
reaction $\pi^-p\rightarrow K^0\pi\Sigma$ seems to give more
weight to the low-energy pole and thus exhibits a peak around
$1390$\,MeV in the $\pi\Sigma$ invariant mass distributions~\cite{Hyodo:2003jw}. Such a two-pole structure of the
$\Lambda(1405)$ has recently been tested by the reaction
$K^-p\rightarrow\pi^0\pi^0\Sigma^0$~\cite{Prakhov:2004an}, as demonstrated in
Ref.~\cite{Magas:2005vu}.

The electromagnetic transition rates of excited baryons to their respective
ground states provide a relatively clean probe of the structure of the baryons.
In this respect, we expect that the radiative decay widths of the $\Lambda(1405)$
can offer us some clues on its two-pole structure. However, the radiative
decays of the excited hyperon states have very small branching ratios and to
date very few electromagnetic transition rates have been measured. Recently,
the CLAS collaboration at Jefferson Lab has reported a new measurement of the
radiative decay widths of the $\Sigma^0(1385)$ and the
$\Lambda(1520)$~\cite{Taylor:2005zw}, which, as argued in
Ref.~\cite{Myhrer:2006hx}, suggests that the wave functions of the hyperon
ground states should contain sizable components of excited quark states
(configuration mixing). As for the $\Lambda(1405)$, there is no direct
measurement of its radiative decay widths. Using an isobar
model  to fit the $K^-p$ atom data of Ref.~\cite{Whitehouse:1989yi}, H. Burkhardt and J. Lowe ~\cite{Burkhardt:1991ms}
obtained the following numbers:
\begin{eqnarray*}
&&\Gamma_{\Lambda(1405)\rightarrow\gamma\Lambda(1116)}=27\pm8\,\mbox{keV},\\
&&
\Gamma_{\Lambda(1405)\rightarrow\gamma\Sigma^0(1193)}=10\pm4\,\mbox{keV}\quad\mbox{or}\quad
23\pm7\,\mbox{keV},
\end{eqnarray*}
which are sometimes quoted as ``experimental data'' in the literature.

On the other hand, as shown in Ref.~\cite{Lee:1998gt}, the U$\chi$PT
model of Refs.~\cite{Oset:1997it,Oset:2001cn} can also describe
rather well the $K^-p$ atom data of Ref.~\cite{Whitehouse:1989yi}.
Therefore, it is truly desirable to calculate the radiative decay
width of the $\Lambda(1405)$ within the same framework. This is the
main purpose of the present work.

This paper is organized as follows. In Sect.~\ref{sec:cc} we give a
brief description of the chiral unitary coupled channel approach. In
Sect.~\ref{sec:decay} we calculate the radiative decay widths of the
$\Lambda(1405)$, discuss how these numbers are closely related to
the chiral structure of our approach and compare our predictions
with those of other theoretical models. In Sect.~\ref{sec:reaction}
we study the reactions $K^-p\rightarrow
\pi^0\gamma\Lambda(\Sigma^0)$ and $\pi^-p\rightarrow K^0\gamma
\Lambda(\Sigma^0)$. There we show that while the first reaction
stresses the high-energy pole of the $\Lambda(1405)$ in both the
$\gamma\Lambda$ and $\gamma\Sigma^0$ channels, the second reaction
gives more weight to the low-energy pole in the $\gamma\Sigma^0$
channel.  Conclusions and a brief summary are presented in
Sect.~\ref{sec:summary}.

\section{Brief description of the two $\Lambda(1405)$ states in
the chiral unitary coupled channel approach\label{sec:cc}}
 In \cite{Oset:1997it,Oset:2001cn,Oller:2000fj,Jido:2003cb},
 the unitary formalism with coupled channels
 using chiral Lagrangian is exposed. The lowest order chiral
 Lagrangian for the interaction of the pseudoscalar mesons of the SU(3)
 octet of the pion with the baryons of the proton octet is used. By
 picking the terms that contribute to the $MB\rightarrow MB$
 amplitude the Lagrangian is given by \cite{Oset:1997it}:
\begin{equation}
\mathcal{L}=\frac{1}{4f^2}\langle\bar{B}i\gamma^\mu[\Phi\partial_\mu\Phi-\partial_\mu\Phi
\Phi,B]\rangle,
\end{equation}
which, projected over $s$-wave, provides tree level transition
amplitudes \cite{Oset:2001cn}:
\begin{eqnarray}
V_{ij}&=&-C_{ij}\frac{1}{4f^2}(2\sqrt{s}-M_{B_i}-M_{B_j})\nonumber\\
&&\quad\quad\quad\times\left(\frac{M_{B_i}+E}{2M_{B_i}}\right)^{1/2}
\left(\frac{M_{B_j}+E'}{2M_{B_j}}\right)^{1/2},
\end{eqnarray}
with $E$, $E'$ ($M_B$) the energies (masses) of the baryons and
$C_{ij}$ coefficients tabulated in \cite{Oset:1997it}. These tree
level amplitudes are used as kernel of the Bethe Salpeter equation
in coupled channels
\begin{equation}
T=[1-Vg]^{-1}V,
\end{equation}
where $V$ appears factorized on shell
\cite{Oset:1997it,Oller:2000fj} and $g$ is the loop function of a
meson and a baryon propagators, regularized by a cut off in
\cite{Oset:1997it} and in dimensional regularization in
\cite{Oller:2000fj,Oset:2001cn,Jido:2003cb}.

For the particular case of ${1/2}^-$ states (in $MB$ $s$-wave
interaction) with strangeness $S=-1$ and zero charge we have ten
channels: $K^-p$, $\bar{K}^0n$, $\pi^0\Lambda$, $\pi^0\Sigma^0$,
$\eta\Lambda$, $\eta\Sigma^0$, $\pi^+\Sigma^-$, $\pi^-\Sigma^+$,
$K^+\Xi^-$, and $K^0\Xi^0$. The explicit solution of the Bethe
Salpeter equation leads to poles in the second Riemann sheet
corresponding to resonances. In this sector one finds two poles
close to the nominal $\Lambda(1405)$ resonance, and other poles
corresponding to the $\Lambda(1670)$ and other $\Sigma$
resonances~\cite{Oset:2001cn,Jido:2003cb}. The pole position
provides the mass and half width (through its imaginary part) and
the residues at the pole give the couplings of the resonance to the
different channels. These couplings will be needed in what follows
to determine the radiative decay widths of the $\Lambda(1405)$. Only
one loop function involving these latter couplings will be used, but
one has to keep in mind that the resonance couplings used in the
evaluation summarize the effect of the multichannel multiple
scattering of the different states prior to the final coupling to
the photon. Although arguments of gauge invariance require the
coupling of the photon to all internal loops of the diagrammatic
series of the Bethe Salpeter equation~\cite{Borasoy:2005zg}, such
loops involve an $s$-wave and a $p$-wave vertex and vanish in the
present case in the large baryon mass limit. In practice, they are
negligible for finite masses \cite{Doring:2006ub,Doring:2005bx}.

In Table \ref{table:jido} we summarize the pole position and
couplings of the two $\Lambda(1405)$ states to the different
channels. These will be used in the next section. We omit in the
table the neutral channels $\eta\Lambda$, $\eta\Sigma$,  and
$\pi\Lambda$, which do not contribute to the radiative decay of the
$\Lambda(1405)$.

\begin{table}[htpb]
\renewcommand{\arraystretch}{1.5}
\setlength{\tabcolsep}{3mm}
 \centering \caption{\label{table:jido}The two poles $z_R$ of the $\Lambda(1405)$ and the corresponding
 couplings to different coupled channels. Taken from Ref.~\cite{Jido:2003cb}.} 
\begin{tabular}{ccccc}
\hline\hline
 $z_R$ & \multicolumn{2}{c}{$(1390-66i)$} & \multicolumn{2}{c}{$(1426-16i)$}\\

    & $g_i$ & $|g_i|$ & $g_i$ & $|g_i|$ \\
    \hline
$\pi\Sigma$ & $-2.5+1.5i $ & 2.9 & $0.42+1.4i$ & 1.5\\
$\bar{K}N$ & $1.2-1.7i$ & 2.1 & $-2.5-0.94i$ & 2.7 \\
$K\Xi$ & $-0.45+0.41i$ & 0.61 & $0.11+0.33i$ & 0.35    \\
 \hline\hline
\end{tabular}
\end{table}

\begin{figure*}[htpb]
 \centering
\includegraphics[scale=0.7]{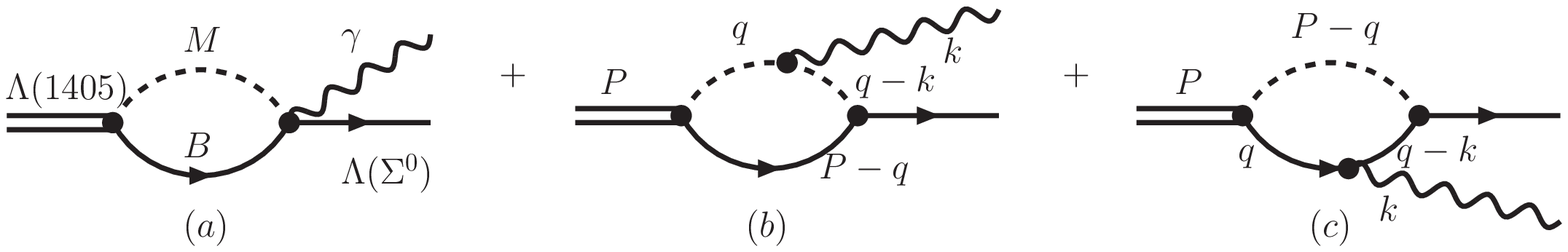}
\caption{\label{figure_decayw}The radiative decay mechanism of the
$\Lambda(1405)$, where $MB$ can be any of the four charged channels
of the ten coupled channels:
 $K^-p$, $\bar{K}^0n$,
$\pi^0\Lambda$, $\pi^0\Sigma^0$, $\eta\Lambda$, $\eta\Sigma^0$,
$\pi^+\Sigma^-$, $\pi^-\Sigma^+$, $K^+\Xi^-$, and $K^0\Xi^0$.}
\end{figure*}

 The model of Ref.~\cite{Oset:1997it} was calibrated
 using the following threshold branching ratios:
\begin{eqnarray*}
    \gamma&=&\frac{\Gamma(K^-p\rightarrow\pi^+\Sigma^-)}{\Gamma(K^-p\rightarrow\pi^-\Sigma^+)}
    =2.36\pm0.04,\\
    R_c&=&\frac{\Gamma(K^-p\rightarrow\mbox{charged
    particles})}{\Gamma(K^-p\rightarrow\mbox{all})} =0.664\pm0.011,\\
    R_n&=&
    \frac{\Gamma(K^-p\rightarrow\pi^0\Lambda)}{\Gamma(K^-p\rightarrow\mbox{all
    neutral states})} =0.189\pm0.015.
\end{eqnarray*}

With the same set of parameters, in Ref.~\cite{Lee:1998gt}, the following
branching ratios are obtained:
\begin{eqnarray*} B_{K^-p\rightarrow\gamma\Lambda}&=&1.10\times10^{-3}
    \,(0.86\pm0.16\times10^{-3}),\\
    B_{K^-p\rightarrow\gamma\Sigma^0}&=&1.05\times10^{-3}\,(1.44\pm0.31\times10^{-3}),\\
        R&=&\frac{B_{K^-p\rightarrow\gamma\Lambda}}{B_{K^-p\rightarrow\gamma\Sigma^0}}
        =1.04 \,(0.4\sim0.9),
\end{eqnarray*}
which are all in reasonable agreement with the data (shown in the
parentheses)~\cite{Whitehouse:1989yi}. We have made use of this model~\cite{Oset:1997it} and have
evaluated the branching ratios for other reactions as shown below:
\begin{eqnarray*}
R_{\pi^0\Lambda}&=&\frac{\Gamma(K^-p\rightarrow\pi^0\Lambda)}{\Gamma(K^-p\rightarrow\mbox{all})}
=0.083\,(0.075),\\
R_{\pi^0\Sigma^0}&=&\frac{\Gamma(K^-p\rightarrow\pi^0\Sigma^0)}{\Gamma(K^-p\rightarrow\mbox{all})}
=0.293\,(0.261),\\
R_{\pi^+\Sigma^-}&=&\frac{\Gamma(K^-p\rightarrow\pi^+\Sigma^-)}{\Gamma(K^-p\rightarrow\mbox{all})}
=0.437\,(0.467),\\
R_{\pi^-\Sigma^+}&=&\frac{\Gamma(K^-p\rightarrow\pi^-\Sigma^+)}{\Gamma(K^-p\rightarrow\mbox{all})}
=0.188\,(0.196),
\end{eqnarray*}
where the numbers in the parentheses are the experimental
data~\cite{Whitehouse:phd}.  These same data
are used to fix the strong coupling constants in the isobar model
of Ref.~\cite{Burkhardt:1991ms} to deduce the radiative decay widths of the $\Lambda(1405)$.

As we have seen, with the use of only one cut-off parameter and the lowest-order chiral
Lagrangian, one can reproduce fairly well all the low energy data related to the $\Lambda(1405)$,
both strong and electromagnetic.  In this work, we extend the unitary
coupled-channel chiral approach of Refs.~\cite{Oset:1997it,Oset:2001cn} to
investigate the radiative decay widths of the $\Lambda(1405)$. We also study
several related reactions to investigate the possibility of experimentally
testing the two-pole structure of the $\Lambda(1405)$ and the predictions of
the present work.

\section{The radiative decay width of the $\Lambda(1405)$}\label{sec:decay}
In the picture of the $\Lambda(1405)$ as a dynamically generated
resonance from the meson baryon interaction, the coupling of the
photon to the resonance proceeds via the coupling to its meson and
baryon components. As mentioned in the former section, gauge
invariance is preserved in this picture as shown in
Ref.~\cite{Borasoy:2005zg}. In practical terms, as shown in
Refs.~\cite{Doring:2006ub,Doring:2007rz}, it means that the
mechanisms for the $\Lambda(1405)$ decay into $\gamma\Lambda$ or
$\gamma\Sigma^0$ are given by the diagrams shown in
Fig.~\ref{figure_decayw}. The corresponding $t$-matrix elements read
\begin{eqnarray}\label{equation:effective_coupling}
-it&=& \sum_i g_{\Lambda(1405)\rightarrow i}(-e)Q_i
\left[\alpha_i\frac{D+F}{2f}+\beta_i\frac{D-F}{2f}\right]\nonumber\\
&&\times G^{\mu\nu}\sigma_\mu\epsilon_\nu\nonumber\\
&\equiv&-ig_{\Lambda(1405)\rightarrow\gamma Y}\vec{\sigma}\cdot\vec{\epsilon}
\quad\mbox{with}\quad Y=\Lambda\,\mbox{or}\,\Sigma^0
\end{eqnarray}
with $\sigma^\mu=(0,\vec{\sigma})$, where $i$ denotes any of the ten coupled channels $K^-p$, $\bar{K}^0n$,
$\pi^0\Lambda$, $\pi^0\Sigma^0$, $\eta\Lambda$, $\eta\Sigma^0$,
$\pi^+\Sigma^-$, $\pi^-\Sigma^+$, $K^+\Xi^-$, and $K^0\Xi^0$.
In Eq.~(\ref{equation:effective_coupling}) $Q_i$ is the electric
charge of the meson of channel $i$, which is $-1$, 0, 0, 0, 0, 0, 1, $-1$, 1, and 0,
respectively, for the ten coupled channels with the order given above.
The coupling constants of the $\Lambda(1405)$ to various channels, $g_{\Lambda(1405)\rightarrow i}$,
are given in Table~\ref{table:jido}. It is
to be noted that the couplings tabulated in Table~\ref{table:jido} are for
isospin channels; therefore, appropriate isospin projections are needed when
used in Eq.~(\ref{equation:effective_coupling}).  The coupling constants
$\alpha_i$ and $\beta_i$  for the 4 charged channels $K^-p$, $\pi^+\Sigma^-$,
$\pi^-\Sigma^+$, and $K^+\Xi^-$ are tabulated in Table~\ref{table_kr}.

 \begin{table}[htpb]
     \renewcommand{\arraystretch}{1.5}
 \setlength{\tabcolsep}{3mm}
\caption{\label{table_kr}SU(3) coupling constants defined in Eq.~(\ref{equation:effective_coupling}) for the channels
$K^-p$, $\pi^+\Sigma^-$, $\pi^-\Sigma^+$, and $K^+\Xi^-$. }
\begin{center}
\begin{tabular}{lcccc}
\hline\hline
  & $K^-p$ & $\pi^+\Sigma^-$ & $\pi^-\Sigma^+$& $K^+\Xi^-$ \\
  \hline  $\alpha_{MB\rightarrow\Lambda}$&$-\frac{2}{\sqrt{3}}$&$\frac{1}{\sqrt{3}}$& $
\frac{1}{\sqrt{3}}$ & $\frac{1}{\sqrt{3}}$\\
$\beta_{MB\rightarrow\Lambda}$&$\frac{1}{\sqrt{3}}$&$\frac{1}{\sqrt{3}}$&
$\frac{1}{\sqrt{3}}$ & $-\frac{2}{\sqrt{3}}$\\
$\alpha_{MB\rightarrow\Sigma^0}$&0&$1$& $-1$ & $1$\\
$\beta_{MB\rightarrow\Sigma^0}$&$1$& $-1$ & $1$&0\\
\hline\hline
\end{tabular}
\end{center}
\end{table}

The loop function $G^{\mu\nu}$ can be
easily calculated by employing gauge invariance (see e.g.
Refs.~\cite{Doring:2007rz,Close:1992ay,Oller:1998ia,Marco:1999df,Roca:2006am}).  Since the only external
momenta available in the present process are $P$ (the $\Lambda(1405)$
4-momentum) and $k$ (the photon 4-momentum), the most general amplitude can be
written as 
\begin{equation}
    T=\tilde{T}G^{\mu\nu}\sigma_\mu\epsilon_\nu,
\end{equation}
with
\begin{equation}
G^{\mu\nu}=ag^{\mu\nu}+bP^\mu P^\nu+cP^\mu k^\nu+d k^\mu P^\nu+ek^\mu k^\nu.
\end{equation}
Due to the Lorentz condition $\epsilon_\nu k^\nu=0$, the two terms proportional
to $c$ and $e$ vanish. Furthermore, gauge invariance requires that
$G^{\mu\nu}k_\nu=0$, i.e.
\begin{equation}
 ak^\mu+bP^\mu(P\cdot k)+dk^\mu (P\cdot k)=0,
\end{equation}
which implies that $b=0$ and
\begin{equation}\label{eq:adrelation}
  a=-d (P\cdot k).
\end{equation}
Following these arguments, $T$ only contains the $a$ and $d$ terms.
This  can be further simplified by noting that in the rest frame of
the $\Lambda(1405)$, $\vec{P}=0$, and taking Coulomb gauge for the photon,
$\epsilon^0=0$, only the $a$ term survives. However, the $a$ term can be more
easily computed by employing the relation of Eq.~(\ref{eq:adrelation}). This is
due to the fact that on one hand, the $d$ coefficient is found convergent since, due
to dimensional reasons, it involves two powers of momentum less in the loop
functions than other individual terms, and on the other hand, there are fewer
terms which contribute to the $d$ coefficient.

Immediately, one realizes that the diagram  $(a)$ shown in
Fig.~\ref{figure_decayw} does not contribute to the $d$ term; therefore to
calculate $G^{\mu\nu}$, we only need to calculate the diagrams $(b)$ and $(c)$.
We first look at the $(b)$ diagram. The corresponding $G^{\mu\nu}$ is
explicitly written as
\begin{eqnarray}
 G^{\mu\nu}_{(b)}&=&-i\int\frac{d^4q}{(2\pi)^4}\frac{1}{q^2-m^2}\frac{1}{(q-k)^2-m^2}\\
 &&\hspace{0.5cm}\times
\frac{2M}{(P-q)^2-M^2+i\epsilon}(q-k)^\mu(2q-k)^\nu\nonumber,
\end{eqnarray}
with $m$ the meson mass and $M$ the baryon mass of the corresponding loop.
By employing the Feynman parameterization~\cite{Peskin:1995ev}
\begin{equation}
 \frac{1}{abc}=2\int^1_0dx\,\int^{1-x}_0dz\,\frac{1}{[ax+b(1-x-z)+cz]^3}
\end{equation}
and the following relation~\cite{Peskin:1995ev}
\begin{equation}
 \int d^4q'\frac{1}{(q'^2+s+i\epsilon)^3}=i\frac{\pi^2}{2}\frac{1}{s+i\epsilon},
\end{equation}
one can obtain $G^{\mu\nu}_{(b)}\sigma_\mu\epsilon_\nu$ as
\begin{eqnarray}\label{eq:dterm}
 G^{\mu\nu}_{(b)}\sigma_\mu\epsilon_\nu&=&
 \frac{2M}{(4\pi)^2}\int^1_0dx\,\int^{1-x}_0dz\,\frac{x(1-z)}{s+i\epsilon}(2P\cdot k)\vec{\sigma}\cdot\vec{\epsilon}\nonumber\\
&\equiv&G^b\vec{\sigma}\cdot\vec{\epsilon}
\end{eqnarray}
with
\begin{equation}\label{eq:sterm}
s=-m^2(1-x)+x[P^2(1-x)-M^2-2P\cdot kz].
\end{equation}
In the same way, one can calculate the $G^{\mu\nu}\sigma_\mu\epsilon_\nu$ term
corresponding to  the diagram $(c)$ by simply exchanging $M$ and $m$ in
Eq.~(\ref{eq:sterm}), and replacing $x(1-z)$ by $-xz$ in Eq.~(\ref{eq:dterm}):
\begin{eqnarray}\label{eq:dterm2}
 G^{\mu\nu}_{(c)}\sigma_\mu\epsilon_\nu&=&
 \frac{2M}{(4\pi)^2}\int^1_0dx\,\int^{1-x}_0dz\,\frac{-xz}{s'+i\epsilon}(2P\cdot k)\vec{\sigma}\cdot\vec{\epsilon}\nonumber\\
 &\equiv&G^c\vec{\sigma}\cdot\vec{\epsilon}
\end{eqnarray}
with
\begin{equation}
s'=-M^2(1-x)+x[P^2(1-x)-m^2-2P\cdot kz].
\end{equation}
It should be noted that we have neglected the magnetic term in calculating the
diagram $(c)$, which is small and vanishes in the heavy baryon limit when integrated over the loop momentum since there
are $p$-wave and $s$-wave vertices in the loop. It is
interesting to note that although the diagrams (a), (b), and (c) are all
divergent by themselves, their sum, however, is finite as can be seen from Eqs.~(\ref{eq:dterm}) and (\ref{eq:dterm2}). The
above loop functions corresponding to the diagrams (b) and (c) can be calculated
analytically, and their explicit form can be found in Ref.~\cite{Doring:2007rz},
where the systematic cancellation of the logarithmic divergences is also shown.

\begin{table*}[htpb]
\renewcommand{\arraystretch}{1.5}
\setlength{\tabcolsep}{3mm}
 \centering \caption{\label{table:decaywidth}The radiative decay widths  of the $\Lambda(1405)$
 predicted by different theoretical models, in units of keV. The values denoted by ``U$\chi$PT'' are
 the results obtained in the present study. The widths calculated for the low-energy pole and high-energy
 pole are separated by a comma.} 
\begin{tabular}{c|ccccc}
\hline\hline
 Decay channel & U$\chi$PT& $\chi$QM \cite{Yu:2006sc} & BonnCQM \cite{VanCauteren:2005sm} &
NRQM & RCQM \cite{Warns:1990xi}\\\hline
 $\gamma\Lambda$ & $16.1,\,64.8$ & 168 & 912 & 143 \cite{Darewych:1983yw}, 200, 154 \cite{Kaxiras:1985zv} & 118 \\\
 $\gamma\Sigma^0$& $73.5,\,33.5$ & 103 & 233 & 91 \cite{Darewych:1983yw}, 72, 72 \cite{Kaxiras:1985zv}& 46 \\\hline\hline
Decay channel & MIT bag \cite{Kaxiras:1985zv} & chiral bag \cite{Umino:1992hi} & soliton~\cite{Schat:1994gm}&algebraic model~\cite{Bijker:2000gq}& isobar fit~\cite{Burkhardt:1991ms}\\\hline
 $\gamma\Lambda$ &60, 17 & 75 &44,40 & 116.9&$27\pm8$\\
$\gamma\Sigma^0$ &18, 2.7 & 1.9 &13,17& 155.7&$10\pm4$ or $23\pm7$\\
 \hline\hline
\end{tabular}
\end{table*}

 The radiative decay width of the $\Lambda(1405)$ is calculated according to
\begin{equation}
\Gamma=\frac{1}{\pi}|g_{\Lambda(1405)\rightarrow\gamma Y}|^2 k\frac{M_Y}{M_{\Lambda(1405)}}
\end{equation}
with $Y=\Lambda$ or $\Sigma^0$ and $k$ the center of mass 3-momentum of the
photon in the $\Lambda(1405)$ rest frame. In this work, as in
Refs.~\cite{Oset:1997it,Oset:2001cn,Lee:1998gt}, we use $f=1.15f_\pi$ with $f_\pi=93$\,MeV, $D+F=1.26$, and $D-F=0.33$.

The radiative decay widths are calculated to be
$\Gamma_{\gamma\Lambda}=64.8$\,keV and $\Gamma_{\gamma\Sigma^0}=33.5$\,keV
for the high-energy pole, and
$\Gamma_{\gamma\Lambda}=16.1$\,keV and $\Gamma_{\gamma\Sigma^0}=73.5$\,keV for the low-energy
pole.
These are tabulated in Table \ref{table:decaywidth} together with the
predictions of various other theoretical models, including the chiral quark
model ($\chi$QM)~\cite{Yu:2006sc}, the Bonn constituent quark
model~\cite{VanCauteren:2005sm}, the non-relativistic quark
models~\cite{Darewych:1983yw,Kaxiras:1985zv}, the relativistic constituent
quark model~\cite{Warns:1990xi}, the MIT bag model~\cite{Kaxiras:1985zv}, the
chiral bag model~\cite{Umino:1992hi}, the soliton model~\cite{Schat:1994gm},
the algebraic model~\cite{Bijker:2000gq}, and the isobar model
fit~\cite{Burkhardt:1991ms} to the branching ratios of the radiative decays of
the $K^-p$ atom~\cite{Whitehouse:1989yi}.

\begin{table}
\renewcommand{\arraystretch}{1.5}
\setlength{\tabcolsep}{4mm}
\centering \caption{\label{table:convol}The averaged radiative decay widths of the $\Lambda(1405)$, in units of keV.
See Eq.~(\ref{eq:gamma_ave}) for details.}
\begin{tabular}{c|c|c}
\hline\hline
  Final states & Low-energy pole & High-energy pole \\
   \hline
    $\gamma\Lambda$ & 36.6&74.6 \\
   $\gamma\Sigma^0$ &78.4&31.9\\
 \hline\hline
\end{tabular}
\end{table}

\begin{table}
\renewcommand{\arraystretch}{1.5}
\setlength{\tabcolsep}{3mm}
 \centering \caption{\label{table:reduced_coupling}
 The effective couplings defined in Eq.~(\ref{equation:reduced_coupling}).} 
\begin{tabular}{ccccc}
\hline\hline
 & $K^-p$ & $\pi^+\Sigma^-$ & $\pi^-\Sigma^+$ & $K^+\Xi^-$\\
   \hline
 $g_{MB\rightarrow\gamma\Lambda}$ & $1.26$ & $0.92$ & $-0.92$ &
 $0.35$\\
 $g_{MB\rightarrow\gamma\Sigma^0}$ & $-0.33$ & $0.93$ & $0.93$ &
 $1.26$ \\
 \hline\hline
\end{tabular}
\end{table}

\begin{table}
\renewcommand{\arraystretch}{1.5}
\setlength{\tabcolsep}{3mm}
\centering \caption{\label{table:nominal}The radiative decay widths of the $\Lambda(1405)$
evaluated at the nominal $\Lambda(1405)$ mass, $M=1406.5\,\mbox{MeV}$, in units of keV. ``Low-energy pole''
and ``High-energy pole''
indicate that the coupling constants for the low-energy pole or high-energy pole from Table \ref{table:jido} are used.
} 
\begin{tabular}{c|c|c}
\hline\hline
  Final states & Low-energy pole&High-energy pole\\
   \hline
    $\gamma\Lambda$ &  23.2& 33.8\\
   $\gamma\Sigma^0$ &82.4&26.2\\
 \hline\hline
\end{tabular}
\end{table}

It is interesting to note that our predictions  for
the high-energy pole seem to agree more with the predictions of other
theoretical models, i.e. they all predict  a larger $\gamma\Lambda$ decay width
than the $\gamma\Sigma^0$ decay width except the algebraic
model~\cite{Bijker:2000gq}. In addition, we note that our predictions for
the high-energy pole are approximately only half of those predicted by the
quark models~\cite{Yu:2006sc,Darewych:1983yw,Kaxiras:1985zv,Warns:1990xi},
which have long been known to fail in describing the $\Lambda(1405)$.

We have studied the effects of the finite width of the $\Lambda(1405)$ on
the calculated decay widths by convoluting
the spectral function of the resonance:
\begin{equation}\label{eq:gamma_ave}
\Gamma_\mathrm{ave.}=\frac{-\frac{1}{\pi}\int\limits^{M+\Gamma}_{M-\Gamma}d\sqrt{s}\,\Gamma_{\gamma
Y}(\sqrt{s})\,\mathrm{Im}\frac{1}{\sqrt{s}-M+i\frac{\Gamma}{2}}
\Theta(\sqrt{s}-\sqrt{s_\mathrm{th}})}{-\frac{1}{\pi}\int\limits^{M+\Gamma}_{M-\Gamma}d\sqrt{s}
\,\mathrm{Im}\frac{1}{\sqrt{s}-M+i\frac{\Gamma}{2}}
\Theta(\sqrt{s}-\sqrt{s_\mathrm{th}})},
\end{equation}
where $M$ and $\Gamma$ are the pole mass and the corresponding width for either of the two poles
of the $\Lambda(1405)$, and $s_\mathrm{th}$ is the threshold of the main decay channel $\pi\Sigma$.
The results are listed in Table \ref{table:convol}. It is easily seen that
they are qualitatively similar to those listed in Table \ref{table:decaywidth},
but the $\gamma\Lambda$ rate for the low-energy pole
has almost doubled, which might indicate relatively large uncertainties in this quantity.

It is instructive to see
the origin of these results. Since the
high-energy pole of the $\Lambda(1405)$ couples more strongly to the $\bar{K}N$
channel and the low-energy pole couples more strongly to the $\pi\Sigma$
channel (see Table \ref{table:jido}), the difference between the results for the high-energy pole and low-energy
pole can be easily understood by noting the chiral structure
of the effective coupling constants $g_{\Lambda(1405)\rightarrow\gamma\Lambda}$
and
$g_{\Lambda(1405)\rightarrow\gamma\Sigma^0}$, see Eq.~(\ref{equation:effective_coupling}).
Neglecting the dependence on the loop functions, the couplings of
$MB\rightarrow\gamma Y$ are proportional to
\begin{equation}\label{equation:reduced_coupling}
    g_{MB\rightarrow\gamma Y}\equiv Q_M\Big[\alpha_{MB\rightarrow Y}(D+F)+\beta_{MB\rightarrow Y}(D-F)\Big]
\end{equation}
where $Y$ is either $\Lambda$ or $\Sigma^0$, and $Q_M$ is the electric charge of the meson. The corresponding
couplings are tabulated in Table \ref{table:reduced_coupling}. From
this table, one immediately realizes that in the decay to the
$\gamma\Lambda$ final state, the contributions of the two intermediate channels
$\pi^+\Sigma^-$ and $\pi^-\Sigma^+$ cancel each other, though not
completely since the corresponding loop functions in these two
channels will differ by a small amount considering that they have
different but quite similar masses.

\begin{figure*}[htpb] \centering
    \includegraphics[scale=0.40]{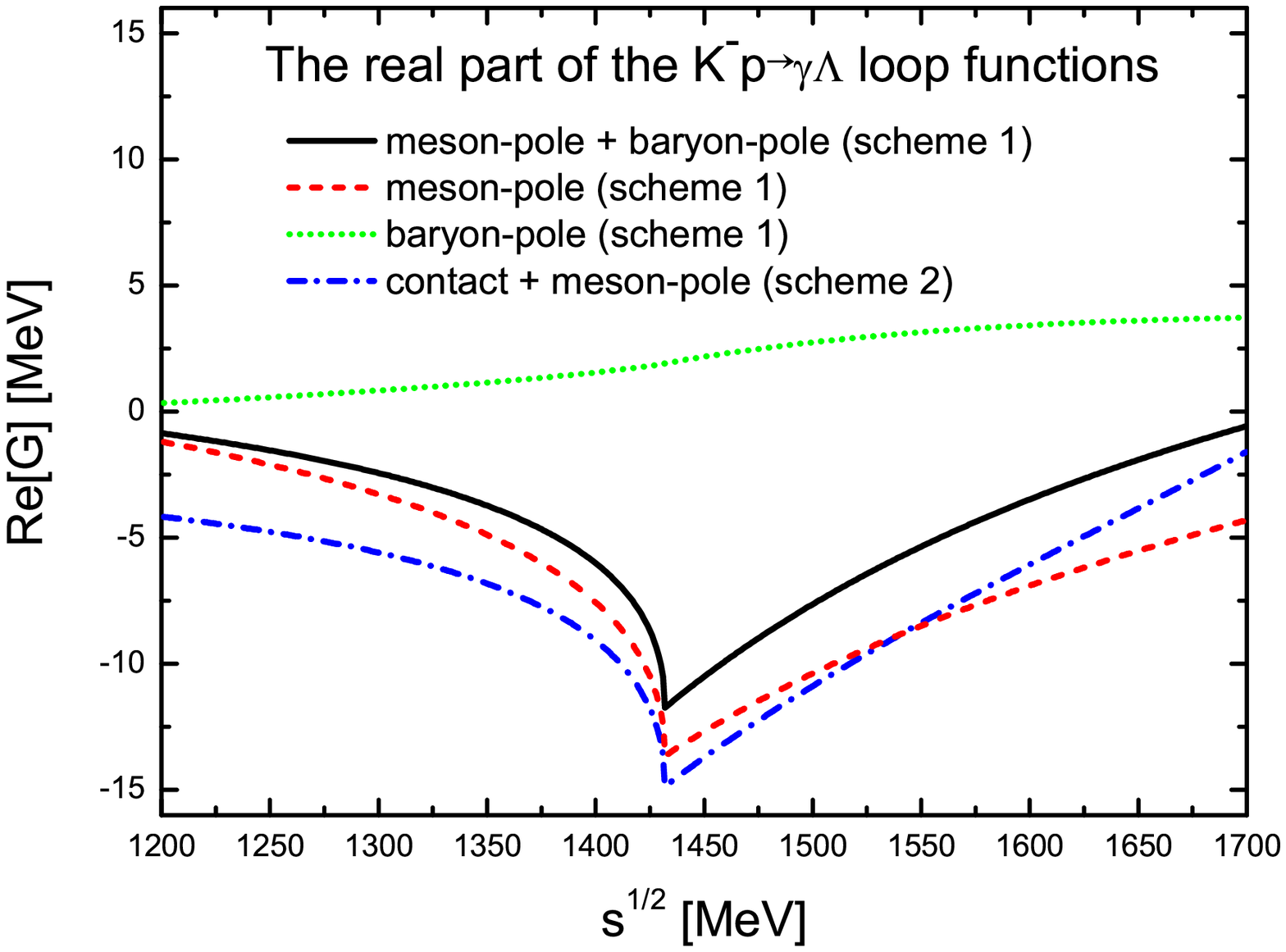}%
    \includegraphics[scale=0.40]{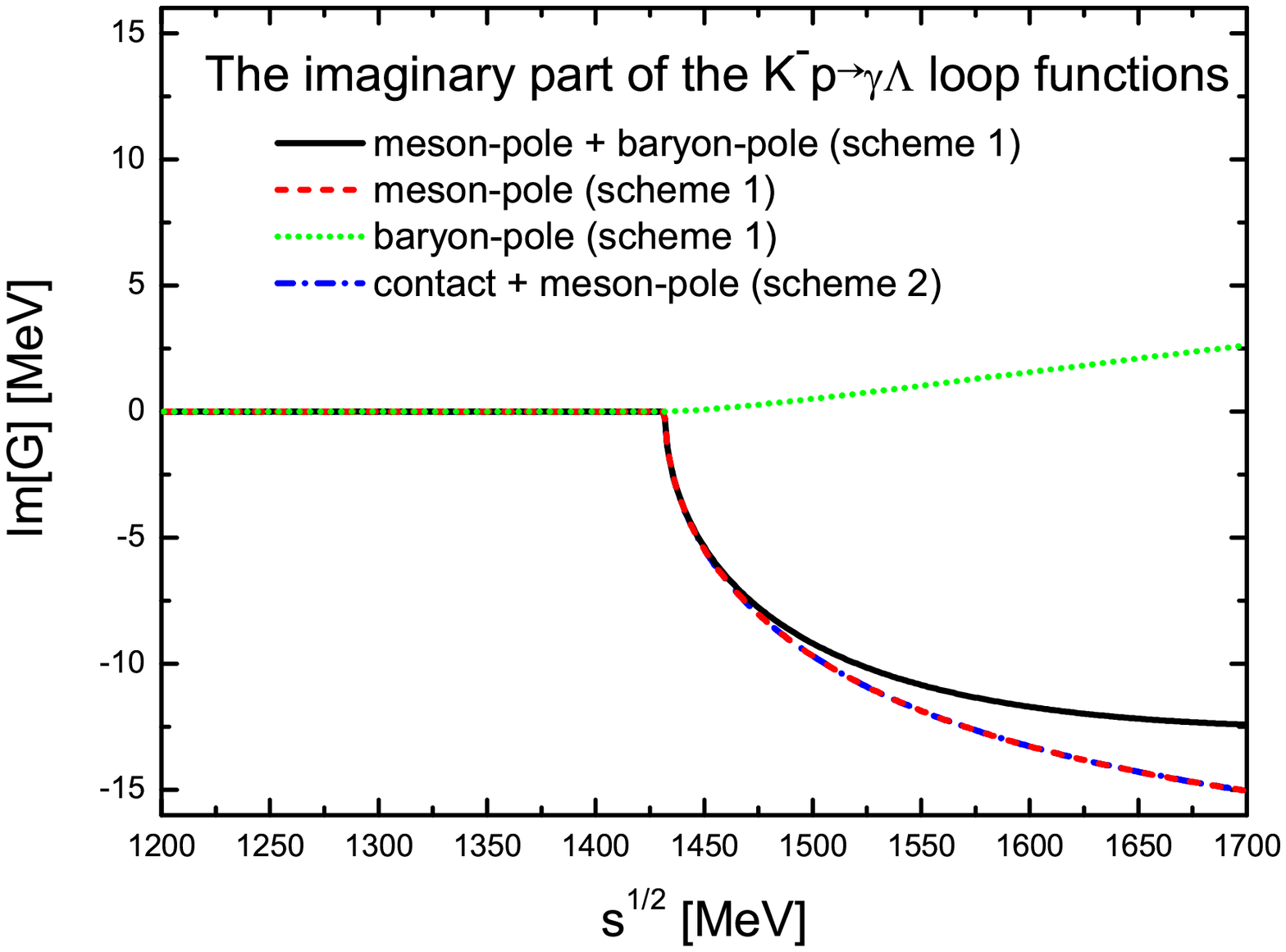}
    \caption{\label{figure_loop}Comparison of the $K^-p\rightarrow
    \gamma\Lambda$ loop functions
     obtained from different schemes. See text for details.}
\end{figure*}
One can qualitatively understand the results for the partial decay widths of the two poles as follows.
The
coupling constants for these two poles differ in the relative
strength of different channels: For the high-energy pole, the coupling to
the $\bar{K}N$ intermediate channel is larger while for the low-energy pole the coupling
to the $\pi\Sigma$ channel is larger. Therefore, for the $\gamma\Lambda$ channel, using the
coupling constants for the low-energy pole instead of the high-energy pole
effectively reduces the contribution of the $K^-p$ channel while
it enhances the contribution of the $\pi^+\Sigma^-$ and $\pi^-\Sigma^+$,
but the contributions of the $\pi^+\Sigma^-$ and $\pi^-\Sigma^+$ almost
cancel each other, and thus, the net effect of using the coupling
constants for the low-energy pole instead of the high-energy pole reduces the
corresponding decay width.

On the other hand, for the decay mode to the $\gamma\Sigma^0$ final state, two
things are noteworthy. First, the contribution of the $K^-p$ channel is much
smaller compared to the contribution of the $K^-p$ channel to the decay mode
$\gamma\Lambda$. Second, the contributions of $\pi^+\Sigma^-$ and
$\pi^-\Sigma^+$ add constructively instead of destructively.
Therefore, when using the coupling constants for the low-energy pole instead of
the high-energy pole, the radiative decay width increases.

At this point, we can conclude that if different experiments actually measure
different radiative decay widths for the $\Lambda(1405)$, it can be used as
evidence for supporting the two-pole structure of the $\Lambda(1405)$.
Instead of the individual radiative decay widths, the ratio between the radiative
decay widths to the $\gamma\Lambda$ and $\gamma\Sigma^0$ final states  might
server better this purpose since in one case one has
$\Gamma_{\gamma\Lambda}/\Gamma_{\gamma\Sigma^0}>1$ and in the other case one
has $\Gamma_{\gamma\Lambda}/\Gamma_{\gamma\Sigma^0}<1$.  This controversy, if
confirmed by experiment, can only be explained by assuming that there are
actually two poles related to the nominal $\Lambda(1405)$.

At first sight, our calculated decay widths are somehow different from the
isobar model fit of H. Burkhardt and J. Lowe~\cite{Burkhardt:1991ms}. However,
one should remember that in their fit they used the nominal $\Lambda(1405)$
mass. As we can see in Table \ref{table:nominal},
when calculated at the nominal $\Lambda(1405)$ mass, with the coupling
constants of the high-energy pole, our calculated radiative decay width for the
$\gamma\Lambda$ channel is 33.8\,keV  and for the $\gamma\Sigma^0$ is 26.2\,keV.
On the other hand, if we use the coupling constants for the low-energy pole,
the results would be 23.2\,keV for the $\gamma\Lambda$ channel and 82.4\,keV
for the $\gamma\Sigma^0$ channel. It is evident that our results with the
high-energy pole coupling constants are in good agreement with the results of
the isobar model
fit: $\Gamma_{\gamma\Lambda}=27\pm8$\,keV and $\Gamma_{\gamma\Sigma^0}=23\pm
7$\,keV or $10\pm4$\,keV~\cite{Burkhardt:1991ms}, if the calculations are done
for the nominal $\Lambda(1405)$ mass. This might indicate that the $K^-p$ intermediate
channel, and thus the high-energy pole, is dominant in the process analyzed in
Ref.~\cite{Burkhardt:1991ms} to deduce the radiative decay widths.

Finally, we would like to stress the importance of the baryon-pole term of the
diagram (c) of Fig.~\ref{figure_decayw}. If instead of employing gauge invariance and calculating all
the three diagrams (a), (b), and (c) of Fig.~\ref{figure_decayw} (we call this
scheme 1), we had calculated only diagrams (a) and (b), i.e. only the contact
term and the meson-pole term (we call this scheme 2), we would have obtained different results.
The corresponding $K^-p\rightarrow \gamma\Lambda$ loop
functions are plotted in Fig.~\ref{figure_loop} as a function of the invariant mass
of the $MB$ system. The
loop functions of scheme 2 are calculated using the cutoff $\Lambda=630$\,MeV as in
Refs.~\cite{Oset:1997it,Lee:1998gt}.
It is easily seen the baryon-pole term changes the real part of the loop
function by $\sim$30\%.  As a consequence, without this term, the calculated radiative
decay width would be larger by almost 40\%.  It is also interesting to note that
the imaginary part of the contact plus meson-pole term of scheme 2 is identical
to the imaginary part of the meson-pole term of scheme 1.

\section{Exploration of possible reactions}\label{sec:reaction}

In the previous section, we have shown that the radiative decay widths of the
$\Lambda(1405)$ to $\gamma\Lambda$ and $\gamma\Sigma^0$ can be very
different depending on which of the two poles dominates.  Therefore, the
good test would be to select different reactions which give different weights to the two
poles. Such reactions can provide further evidence for the
predicted two-pole structure of the $\Lambda(1405)$.  A first evidence for the
two-pole structure of the $\Lambda(1405)$ has been provided in
Ref.~\cite{Magas:2005vu}. By investigating the invariant mass distribution of
the $\pi^0\Sigma^0$ final state in the $K^-p\rightarrow\pi^0\pi^0\Sigma^0$
reaction, it was shown that this reaction gives more weight to the high-energy
pole of the $\Lambda(1405)$ in contrast to the reaction $\pi^-p\rightarrow
K^0\pi\Sigma$, where the low-energy pole is believed to be dominant. This
$\pi^-p\rightarrow K^0\pi\Sigma$ reaction has been studied in
Ref.~\cite{Hyodo:2003jw}. The authors found that the pure chiral mechanism
alone cannot reproduce the experimental invariant mass distributions. However,
by explicitly taking into account the contribution of the $N^*(1710)$, which
gives more weight to the $\pi\Sigma$ channel, and thus more weight to the low-energy
pole, they found that the experimental data can be reasonably described.  In
the following, we study the corresponding electromagnetic reactions
$K^-p\rightarrow \pi^0\gamma \Lambda(\Sigma^0)$ and $\pi^- p\rightarrow
K^0\gamma \Lambda(\Sigma^0)$ in more detail.
\begin{figure*}[htpb]
 \centering
\includegraphics[scale=0.7]{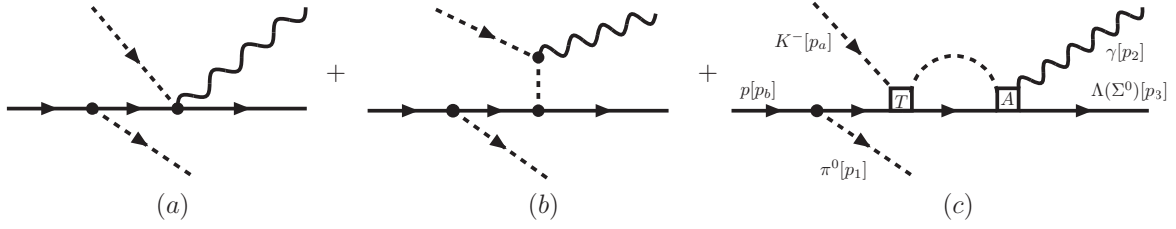}
\caption{\label{figure_kp}The diagrams contributing to the reaction $K^-p\rightarrow \pi^0\gamma\Lambda(\Sigma^0)$.}
\end{figure*}
\begin{figure*}[htpb]
 \centering
 \includegraphics[scale=0.7]{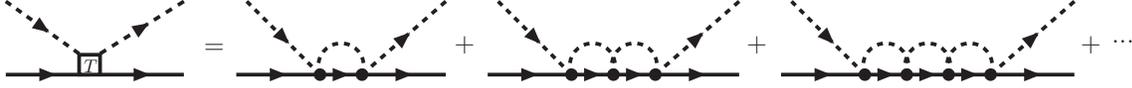}
 \caption{\label{figure:tmatrix}The strong amplitude $T$.}
\end{figure*}
\begin{figure*}[htpb]
 \centering
\includegraphics[scale=0.7]{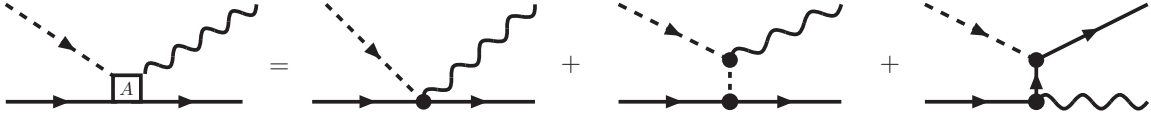}
\caption{\label{figure:gmatrix}The electromagnetic amplitude $A$.}
\end{figure*}

\subsection{The $K^-p\rightarrow \pi^0\gamma\Lambda(\Sigma^0)$ reaction}

In Ref.~\cite{Magas:2005vu}, it was found that the reaction $K^-p\rightarrow
\pi^0\pi^0\Sigma^0$ is dominated by the so-called nucleon-pole mechanism. By
analogy, the $K^-p\rightarrow \pi^0\gamma \Lambda(\Sigma^0)$ reaction should
also proceed through the same mechanism, which is explicitly shown in
Figs.~\ref{figure_kp}, \ref{figure:tmatrix}, and \ref{figure:gmatrix}.  The corresponding $t$-matrix reads
\begin{equation}
-it=-iA\vec{\sigma}\cdot\vec{\epsilon}\,\vec{\sigma}\cdot\vec{v}-iB\vec{\epsilon}\cdot\vec{p}_a\,\vec{\sigma}\cdot
(\vec{p}_a-\vec{p}_2)\,\vec{\sigma}\cdot\vec{v},
\end{equation}
where $\vec{v}=\vec{p}_1(1+\frac{p_1^0}{2M_N})+\frac{p_1^0}{M_N}\vec{p}_a$, $A$ and $B$ are
\begin{eqnarray}\label{eq:tmatrixA}
A&=&\frac{D+F}{2f_\pi}\frac{M_N}{E_N(\vec{p}_a+\vec{p}_1)}\frac{1}{E_N-p_1^0-E_N(\vec{p}_a+\vec{p}_1)}\nonumber\\
&&\times
\Bigg[\sum_j T_{k^-p\rightarrow j}(M_I)(G^b(M_I)+G^c(M_I))\nonumber\\
&&\hspace{1.2cm}\times(-e)Q_j\left(\alpha_j\frac{D+F}{2f}+\beta_j\frac{D-F}{2f}\right)\nonumber\\
&&\hspace{0.6cm}
+(-e)Q_1\left(\alpha_1\frac{D+F}{2f}+\beta_1\frac{D-F}{2f}\right)\Bigg],
\end{eqnarray}
\begin{eqnarray}
B&=&\frac{D+F}{2f_\pi}\frac{M_N}{E_N(\vec{p}_a+\vec{p}_1)}\frac{1}{E_N-p_1^0-E_N(\vec{p}_a+\vec{p}_1)}\nonumber\\
&&\times\frac{-2eQ_1}{(p_a-p_2)^2-m_a^2}\left[\alpha_1\frac{D+F}{2f}+\beta_1\frac{D-F}{2f}\right],
\end{eqnarray}
with $M_N$ the nucleon mass, $E_N$ the nucleon energy, and $M_I^2=(p_2+p_3)^2$. The last term of Eq.~(\ref{eq:tmatrixA}) accounts for the first tree
level diagram of Fig.~\ref{figure_kp}, while the other part of the $A$ coefficient proportional to
$T_{K^-p\rightarrow j}$ accounts for the third diagram (loop diagram) of Fig.~\ref{figure_kp}. The term $B$
corresponds to the second tree level diagram of Fig.~\ref{figure_kp}.
In Eq.~(\ref{eq:tmatrixA}), the $T_{k^-p\rightarrow j}$, with $j$ referring to the
ten coupled channels, are the
strong amplitudes of Ref.~\cite{Oset:2001cn}, which is diagrammatically shown in Fig.~\ref{figure:tmatrix}. The loop functions $G^b$ and $G^c$ are those of Eqs.~(\ref{eq:dterm}) and (\ref{eq:dterm2}).

\begin{figure}[tp]
 \centering
\includegraphics[scale=0.42]{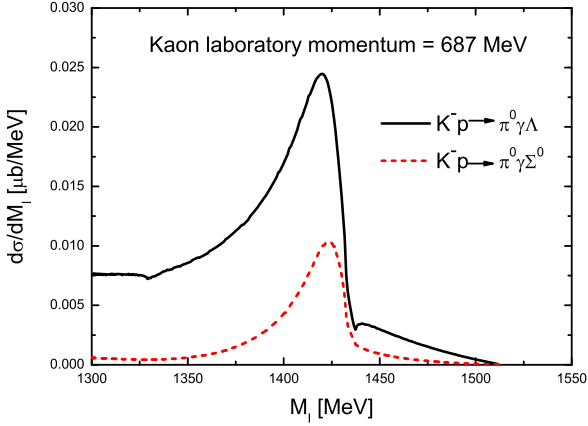}
\caption{\label{figure_kp2}The invariant mass distribution of $K^-p\rightarrow \pi^0\gamma
\Lambda(\gamma\Sigma^0)$ as a function of the invariant mass of the final
$\gamma\Lambda$($\gamma\Sigma^0$) system.}
\end{figure}

The invariant mass distribution is then calculated by
\begin{eqnarray}\label{eq:inv:infra}
\frac{d\sigma}{dM_I}&=&\frac{1}{4}\frac{M_N M_Y}{\lambda^{1/2}(s,M_N^2,m_K^2)}\frac{M_I}{\sqrt{s}}\frac{1}{(2\pi)^4}\nonumber\\
&&\times\int^1_{-1}d\cos\theta_1
\int^{E_\mathrm{max}}_{E_\mathrm{min}} dE_2\int^{2\pi}_0d\phi_{12}\bar{\sum}\sum|t|^2\nonumber\\
&&\hspace{1cm}\times\theta(1-\cos^2(\theta_{12}))
\end{eqnarray}
where $\sqrt{s}$ is the center of mass energy of $K^- p$, $\theta_1$ is the angle between $\vec{p}_a$ and $\vec{p}_1$, $\theta_{12}$ is the angle between
$\vec{p}_1$ and $\vec{p}_2$, fixed by kinematics, while $\phi_{12}$ is the azimuthal angle of $\vec{p}_2$
with respect to a frame where $\vec{p}_1$ is chosen in the $z$ direction. In addition,
$E_\mathrm{min}=0$,
$E_\mathrm{max}=\frac{s-(m_1+m_3)^2}{2\sqrt{s}}$, and
\begin{equation}
\cos(\theta_{12})=\frac{1}{2|\vec{p}_1||\vec{p}_2|}\left\{
(\sqrt{s}-E_1-E_2)^2-m^2_3-\vec{p}_1^2-\vec{p}_2^2\right\},
\end{equation}
\begin{eqnarray}
\bar{\sum}\sum|t|^2&=&|v|^2
 \Bigg\{\{2|A|^2+\left(|\vec{p}_a|^2-\frac{(\vec{p}_a\cdot\vec{p}_2)^2}{|\vec{p}_2|^2}\right)\nonumber\\
&&\times\left[(A^*B+AB^*)+
|B|^2(\vec{p}_a-\vec{p}_2)^2\right]\Bigg\}.\nonumber\\
\end{eqnarray}
The invariant mass distributions for the reactions $K^-p\rightarrow \pi^0\gamma\Lambda$ and
$K^-p\rightarrow \pi^0\gamma\Sigma^0$ for a kaon of laboratory momentum 687\,MeV
are shown in Fig.~\ref{figure_kp2}. It is seen
that both the invariant mass distributions exhibit a peak at $\sim$1420\,MeV,
 and therefore manifesting the high-energy pole of the $\Lambda(1405)$.  The invariant
mass distribution is also different from a Breit-Wigner shape, particularly
that of the $\gamma\Lambda$ channel, which is due to the background terms (the
tree-level diagrams in Fig.~\ref{figure_kp}).

It is interesting to recall that in Refs.~\cite{Oset:1997it,Oset:2001cn}, the
two poles,  one at 1390\,MeV with a width of 132\,MeV and the other at
1426\,MeV with a width of 32\,MeV, are generated through the interaction of the ten coupled
channels. The low-energy pole couples more strongly to the $\pi\Sigma$ channel
while the high-energy pole couples more strongly to the $\bar{K}N$ channel.  It
was argued in Ref.~\cite{Jido:2003cb} that reactions favoring different
channels would lead to different invariant mass distributions giving more
weight to one pole or the other.  It should be noted that in
Ref.~\cite{Oset:1997it}, the experimental invariant mass distribution was
produced by the following formula
\begin{equation}\label{eq:inv}
\frac{d\sigma}{dm_\alpha}=C|T_{\pi\Sigma\rightarrow\pi\Sigma}|^2P_\mathrm{CM},
\end{equation}
and thus giving more weight to the low-energy pole and resulting in a peak at
$\sim$1400\,MeV. However, in the presence of two resonances, the different
$T_{ij}$ amplitudes do not peak at the same place, and particularly, $T_{\bar{K}N\rightarrow\pi\Sigma}$
peaks at higher energies than $T_{\pi\Sigma\rightarrow\pi\Sigma}$~\cite{Jido:2003cb}.
In such a case, Eq.~(\ref{eq:inv}) should be replaced ~\cite{Jido:2003cb} by
\begin{equation}
    \frac{d\sigma}{dm_\alpha}=|\sum C_i T_{i\rightarrow\pi\Sigma}|^2P_\mathrm{CM}.
\end{equation}
Hence, if the reaction mechanism does not
completely forbid a $\bar{K}N$ channel, then the peak should always be shifted
towards higher energy.  Thus, it is not surprising that the invariant mass distributions of all the three
reactions~\cite{Nacher:1998mi,Nacher:1999ni,Magas:2005vu} studied previously
including the present one exhibit a peak around $\sim$1420\,MeV.
\begin{figure}
 \centering
    \includegraphics[scale=0.42]{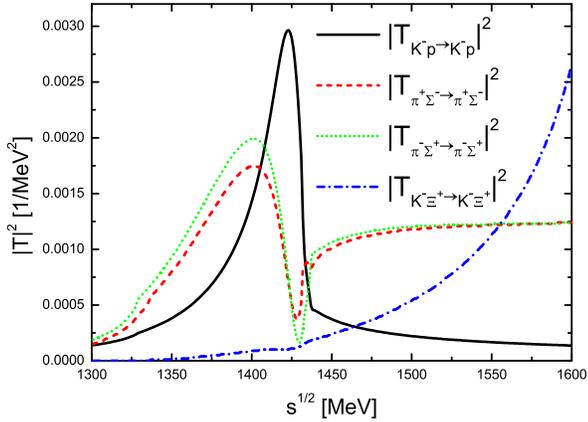}
\caption{The modulus squared of the strong amplitudes obtained from the model of
Ref.~\cite{Oset:2001cn}.}
\label{fig:amplitude}
\end{figure}

It is worth pointing out an interesting difference between the reactions studied in
this work and the other reactions~\cite{Nacher:1998mi,Nacher:1999ni,Magas:2005vu}. In the reaction
$K^-p\rightarrow\pi^0\gamma\Lambda(\Sigma^0)$ and the reaction
$\pi^-p\rightarrow K^0\gamma\Lambda(\Sigma^0)$, which will be studied below, the
$K^-p\rightarrow K^-p$ channel appears as an intermediate channel. Since the
magnitude of the $t$-matrix of this channel is much larger than those of the other channels,
and since
this channel manifests the high-energy pole (see Fig.~\ref{fig:amplitude}), one can always expect a peak at
$\sim1420$\,MeV in the invariant mass distribution of the final states unless
some reaction mechanisms largely suppress this channel. On the other hand, in
the $\pi\Sigma$ final states studied in
Refs.~\cite{Nacher:1998mi,Nacher:1999ni,Magas:2005vu}, the $T_{\bar{K}N\rightarrow\bar{K}N}$
does not
contribute and because the $\bar{K}N\rightarrow \pi\Sigma$ and
$\pi\Sigma\rightarrow\pi\Sigma$ amplitudes have
similar strength (in fact the modulus of the
$K^-p\rightarrow \pi\Sigma$ amplitude
is still approximately two times larger than that of the $\pi\Sigma\rightarrow\pi\Sigma$ amplitude
at their respective peak positions), the invariant mass distributions
will be a superposition of the two peaks, and, depending on the reaction mechanism,
the final distribution will peak at one or another energy.  This situation is somewhat similar to the two-pole
structure of the $K_1(1270)$~\cite{Geng:2006yb}. There, it was found that due to the
dominance of the $\rho K\rightarrow \rho K$ amplitude over the other amplitudes
leading to the $\rho K$ final states, a prominent peak at $\sim 1280\,\mbox{MeV}$ would be
preferred in the invariant mass distribution of the $K\pi\pi$ system leading to
$\rho K$ final states.

\begin{figure*}[htpb]
 \centering
\includegraphics[scale=0.7]{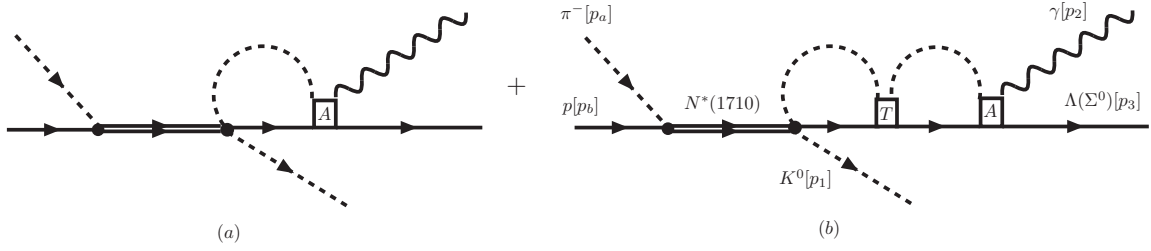}%

\caption{\label{figure_nstar1}The diagrams contributing to the reaction $\pi^-p\rightarrow K^0\gamma\Lambda(\Sigma^0)$.}
\end{figure*}
\subsection{The $\pi^- p\rightarrow K^0\gamma\Lambda(\Sigma^0)$ reaction}

In Ref.~\cite{Hyodo:2003jw}, it was shown that the reaction $\pi^-p\rightarrow K^0\pi\Sigma$ can
be reasonably described in terms of $t$-channel and $s$-channel resonance exchanges.
The corresponding electromagnetic reaction $\pi^-p\rightarrow K^0\gamma\Lambda(\Sigma^0)$, in principle, should
also proceed through similar mechanisms. Nevertheless,
since we only want to make an exploratory study of this reaction, we neglect the
mechanisms of $t$-channel meson exchange (and associated contact $MMBBB$ terms),
referred to as chiral terms in Ref.~\cite{Hyodo:2003jw},  and concentrate on the
$s$-wave $N^*(1710)$ resonance contribution, which was found to be largely dominant in Ref.~\cite{Hyodo:2003jw}.

The $t$-matrix element corresponding to the resonance mechanism shown in Fig.~\ref{figure_nstar1}
reads
\begin{eqnarray}
-it&=&(-it_{\pi^-p\rightarrow N^*})\frac{i}{\sqrt{s}-M_\mathrm{R}+i\frac{\Gamma}{2}}\nonumber\\
&&\hspace{1cm}\times\sum_i(-it_{N^*\rightarrow K^0i})
(-it_{i\rightarrow \gamma\Lambda(\Sigma^0)})
\end{eqnarray}
with
\begin{equation}
-it_{\pi^-p\rightarrow N^*}=\frac{A}{f_\pi}\vec{\sigma}\cdot\vec{p}_a,
\end{equation}
\begin{equation}\label{eq:suppression}
-it_{N^*\rightarrow K^0i}=i\frac{\tilde{B}}{f^2}C_{i}(\omega_{i}-\omega_{K^0}),
\end{equation}
\begin{eqnarray}
-it_{i\rightarrow\gamma\Lambda(\Sigma^0)}&=&
\sum_j\left(G_i(M_I)T_{i\rightarrow j}(M_I)+\delta_{ij}\right)(G^b_{j}+G^c_j) \vec{\sigma}\cdot\vec{\epsilon}\nonumber\\
&&\times
\left[
-ieQ_j\left(\alpha_j\frac{D+F}{2f}+\beta_j\frac{D-F}{2f}\right)\right],
\end{eqnarray}
where $i$, $j$ can be any of the ten coupled channels. The couplings constants $C_i$
can be found in Table II of Ref.~\cite{Hyodo:2003jw}. The loop function $G_i$ is
that of one meson and one baryon, which is calculated in the
dimensional regularization scheme and with the same
subtraction constants as in Ref.~\cite{Oset:2001cn} . The meson energies $\omega_i$ and
$\omega_{K^0}$ are calculated by
\begin{equation}
    \omega_{K^0}=\frac{s+m^2_K-M^2_I}{2\sqrt{s}},
\end{equation}
\begin{equation}
    \omega_i=\frac{M^2_I+m^2_i-M^2_i}{2M_I},
\end{equation}
with $\sqrt{s}$ the invariant mass of $\pi^- p$, $m_i$ and $M_i$ the meson and baryon masses of channel $i$, and $M_I$ the
$\gamma\Lambda(\Sigma^0)$ invariant mass.

\begin{figure}[tpb]
 \centering
\includegraphics[scale=0.42]{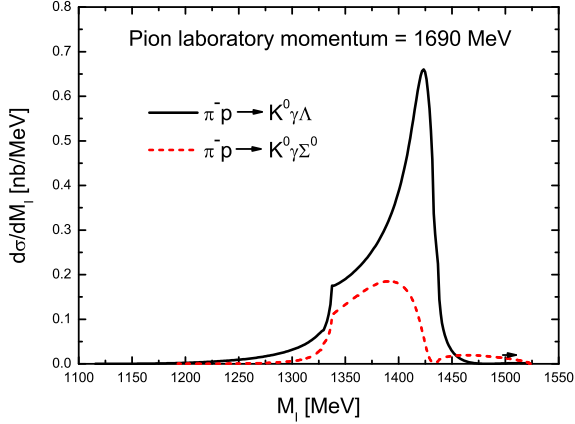}
\caption{\label{figure_nstar2}The invariant mass distribution of $\pi^-p\rightarrow K^0\gamma
\Lambda(\gamma\Sigma^0)$ as a function of the invariant mass of the final $\gamma\Lambda$($\gamma\Sigma^0$) system.}
\end{figure}
In our calculation, the
parameter set II of Ref.~\cite{Hyodo:2003jw} is used, i.e. $M_R=1740$\,MeV,
$|A|=0.1344$, $|\tilde{B}|=0.842$, $\Gamma(\sqrt{s}=2020\mbox{ MeV})=776$\,MeV.
The invariant mass distributions are calculated by
\begin{equation}
\frac{d\sigma}{dM_I}=\frac{M_N M_Y}{\lambda^{1/2}(s,M^2_N,m^2_\pi)}\frac{1}{(2\pi)^3}\frac{1}{\sqrt{s}}p_1\tilde{p}_2
\bar{\sum}\sum|t|^2
\end{equation}
with
\begin{equation}
p_1=\frac{\lambda^{1/2}(s,M_I^2,m_1^2)}{2\sqrt{s}},\quad \tilde{p}_2=\frac{\lambda^{1/2}(M_I^2,m_2^2,m_3^2)}{2M_I},
\end{equation}
and are shown in Fig.~\ref{figure_nstar2} for a pion of laboratory momentum 1690\,MeV.
At first sight, two things are
surprising. First, the $\pi^-p\rightarrow K^0\gamma\Lambda$ reaction still exhibits a
peak around $\sim$1420\,MeV. Second, the magnitude of the $\pi^-p\rightarrow
K^0\gamma\Lambda$ is still larger than that of $\pi^-p\rightarrow
K^0\gamma\Sigma^0$. This is due to the following reasons: As we have
noticed in the previous section, the $\pi^+\Sigma^-\rightarrow
\gamma\Lambda$ term and the $\pi^-\Sigma^+\rightarrow \gamma\Lambda$ term
nearly cancel each other. Therefore, it is natural that the
$\pi^-p\rightarrow K^0\gamma\Lambda$ reaction still manifests the high-energy pole of the
$\Lambda(1405)$ since in this reaction the $K^-p\rightarrow \gamma\Lambda$
intermediate channel contributes most.  While in the $\pi^-p\rightarrow K^0\gamma\Sigma^0$ reaction,
the contribution of $K^-p\rightarrow \gamma\Sigma^0$ itself is small and is
further suppressed by the $\omega_{K^-}-\omega_{K^0}$ factor of Eq.~(\ref{eq:suppression}). On the other
hand, the $\pi^+\Sigma^-\rightarrow\gamma\Sigma^0$ and
$\pi^-\Sigma^+\rightarrow\gamma\Sigma^0$ terms  add constructively and both give
more weight to the low-energy pole of the $\Lambda(1405)$. Therefore, the net
result is a broad peak around $\sim$1390\,MeV, in agreement with the finding of
Ref.~\cite{Hyodo:2003jw}. The larger magnitude of $\pi^-p\rightarrow
K^0\gamma\Lambda$ is due to the fact that the strong amplitude $K^-p\rightarrow
K^-p$ is much larger than the other $i\rightarrow j$ amplitudes, as we have discussed previously.

\section{Summary and conclusions}\label{sec:summary}
Using the unitary extension of the chiral perturbation  theory U$\chi$PT, we
have calculated the radiative decay widths of the $\Lambda(1405)$. Since there
are two poles in the U$\chi$PT
models corresponding to the nominal $\Lambda(1405)$, our calculations, using the model of Refs.~\cite{Oset:1997it,Oset:2001cn}, result in two different
radiative decay widths. For the high-energy pole, our calculated widths,
$\Gamma_{\gamma\Lambda}$=64.8\,keV and $\Gamma_{\gamma\Sigma^0}$=33.5\,keV, are
in qualitative agreement with the predictions of the isobar model fit, in
particular, when evaluated at the nominal $\Lambda(1405)$ mass where the agreement is
very good, but they are in sharp contrast
with those of the quark models and the bag models.  The disagreement with the
quark model predictions adds to the list of other magnitudes that
the quark models also fail to reproduce. For instance, the $\Lambda(1405)$ and the
$\Lambda(1520)$ are degenerate in the model of Ref.~\cite{Isgur:1978xj}.

We also evaluated the radiative decay width for the low-energy pole, and we obtain a totally different
result with $\Gamma_{\gamma\Lambda}=16.1$\,keV and
$\Gamma_{\gamma\Sigma^0}=73.5$\,keV. These are completely different from all
the existing model predictions. All the other theoretical models predict a larger
$\gamma\Lambda$ decay width while a smaller $\gamma\Sigma^0$ decay width except
the  algebraic model~\cite{Bijker:2000gq}.

To find a possible reaction which might give different weights to the two
poles of the $\Lambda(1405)$, we have studied the reactions
$K^-p\rightarrow\pi^0\gamma\Lambda(\Sigma^0)$ and $\pi^-p\rightarrow
K^0\gamma\Lambda(\Sigma^0)$.  These two reactions share a lot of similarities
with the corresponding hadronic reactions, $K^-p\rightarrow\pi^0\pi^0\Sigma^0$
and $\pi^-p\rightarrow K^0\pi\Sigma$, which have been previously studied in
Refs.~\cite{Magas:2005vu,Hyodo:2003jw} and found to yield reasonable agreement
with the data. Our studies show that both these reactions yield a larger
$\gamma\Lambda$ cross section and a smaller $\gamma\Sigma^0$ cross section.
This reflects a non-trivial feature of the U$\chi$PT model: the magnitude of the
$K^-p\rightarrow K^-p$ amplitude is much larger than that of the other amplitudes. On
the other hand, there are subtle differences between the two reactions studied,
$K^-p\rightarrow\pi^0\gamma\Lambda(\Sigma^0)$ and $\pi^- p\rightarrow K^0\gamma\Lambda(\Sigma^0)$.
While
the first reaction gives more weight to the high-energy pole in both channels,
the second reaction gives more weight to the high-energy pole in the
$\gamma\Lambda$ channel and  to the
low-energy pole in the $\gamma\Sigma^0$ channel. This is
reflected by exhibiting a narrower peak at $\sim1420$\,MeV in
the former channel and a broader peak at $\sim1390$\,MeV in the latter channel.
The total cross sections for the $K^-p$ reaction
at $p_K(\mbox{lab})=687$\,MeV  are 1.78\,$\mu b$
($\gamma\Lambda)$ and 0.41\,$\mu b$ ($\gamma\Sigma^0$), which are integrated [see Eq.~(\ref{eq:inv:infra})] with the lower
limit $M_I=1300\,$MeV to avoid infrared divergence. The cross sections
for the $\pi^-p$ reaction
at $p_\pi(\mbox{lab})=1690$\,MeV turn out to be $3.90\times10^{-2}$\,$\mu b$
($\gamma\Lambda$) and $1.58\times 10^{-2}$\,$\mu b$ ($\gamma\Sigma^0$).

Therefore, an experimental measurement of the radiative decay widths of the
$\Lambda(1405)$ in the related reactions, such as
$K^-p\rightarrow\pi^0\gamma\Lambda(\Sigma^0)$ and $\pi^-p\rightarrow K^0\gamma\Lambda(\Sigma^0)$, not only
would lend further support to
the predicted two-pole structure of the $\Lambda(1405)$ but also to the underlying
chiral unitary approach, which so far
has provided a systematic and consistent description of the $\Lambda(1405)$ and low-energy
reactions involving it.

\section{Acknowledgments}
L. S. Geng acknowledges useful communications with Dr. T. Hyodo and financial support
from the Ministerio de Educacion y Ciencia in the Program of estancias de doctores y
tecnologos extranjeros. This work is
partly supported by DGICYT contract number Fis2006-03438 and the Generalitat
Valenciana.
This research is part of the EU Integrated Infrastructure Initiative Hadron
Physics Project under contract number RII3-CT-2004-506078.

\appendix
\section{Basic diagrams}

\begin{enumerate}

\item The lowest-order interaction Lagrangian related to the $MBB$ term is
  \begin{equation}\label{eq:MBB}
  \mathcal{L}=\frac{D+F}{2}\langle \bar{B}\gamma^\mu\gamma_5 u_\mu
  B\rangle+\frac{D-F}{2}\langle \bar{B}\gamma^\mu\gamma_5B
  u_\mu\rangle,
  \end{equation}
where  $u_\mu=-\frac{\sqrt{2}}{f}\partial_\mu\Phi$.
This leads to the following Feymann rule (with incoming meson momentum $k_\mu$):
\begin{equation}
-it=i\mathcal{L}=-\gamma^\mu\gamma_5
k_\mu\left(\alpha\frac{D+F}{2f}+\beta\frac{D-F}{2f}\right).
\end{equation}
With the non-relativistic reduction $\gamma^\mu
\gamma_5 k_\mu\rightarrow -\vec{\sigma}\vec{k}$, the $t$-matrix reads:
\begin{equation}
-it=i\mathcal{L}=\vec{\sigma}\vec{k}\left(\alpha\frac{D+F}{2f}+\beta\frac{D-F}{2f}\right).
\end{equation}

\item The contact (Kroll-Ruderman) term can be obtained by applying the minimal
substitution in Eq.~(\ref{eq:MBB}), i.e. $\partial_\mu\Phi\rightarrow
(\partial_\mu+ieA_\mu)\Phi$, which leads to the following Feynman rule:
  \begin{equation}
 (-it)=i\mathcal{L}=-eQ\vec{\sigma}\vec{\epsilon}
 \left(\alpha\frac{D+F}{2f}+\beta\frac{D-F}{2f}\right),
 \end{equation}
where $Q$ is the charge of the meson.

\end{enumerate}

\end{document}